\documentclass[namedreferences]{SolarPhysics}
\usepackage[optionalrh,solaenum]{spr-sola-addons} 
\usepackage{graphicx}
\usepackage{color}
\usepackage{url}                         % For breaking URLs easily through lines
\usepackage{float}
\begin{document}

\begin{article}

\begin{opening}


\title{Imaging Spectroscopy of a White-Light Solar Flare}


\author{J.C.~\surname{Mart\'inez Oliveros}$^{1}$\sep
S.~\surname{Couvidat}$^{2}$\sep
J.~Schou$^{2}$\sep
S.~\surname{Krucker}$^{1,5}$\sep 
C.~\surname{Lindsey}$^{4}$\sep 
H.S.~\surname{Hudson}$^{1,3}$\sep
P.~\surname{Scherrer}$^{2}$
}

\runningtitle{HMI/RHESSI White Light Flare}
\runningauthor{J.C. Mart\'{\i}nez Oliveros \textit{et al}.}

%% affiliation information with \institute, with one command per each

\institute{
$^{1}$ Space Sciences Laboratory, University of California, Berkeley, CA USA;  email: \url{oliveros@ssl.berkeley.edu}, \url{krucker@ssl.berkeley.edu}, \url{hhudson@ssl.berkeley.edu}\\
$^{2}$ W.W.Hansen Experimental Physics Laboratory, Stanford University, Stanford, CA USA;
email: \url{schou@sun.stanford.edu}, \url{pscherrer@solar.stanford.edu}\\
$^{3}$ Department of Physics \& Astronomy, University of Glasgow, Glasgow, Scotland, UK\\
$^{4}$ North West Research Associates, CORA Division, Boulder, CO USA;
email: \url{clindsey@cora.nwra.com}\\
$^{5}$ University of Applied Sciences North Western Switzerland, Institute of 4D Technologies, 5210 Windisch, Switzerland
}


\begin{abstract}

We report observations of a white-light solar flare ({\tt  SOL2010-06-12T00:57}, M2.0) observed by the 
\textit{Helioseismic Magnetic Imager} (HMI) on the \textit{Solar Dynamics Observatory} (SDO) and the \textit{Reuven Ramaty High-Energy Solar Spectroscopic Imager} (RHESSI).
The HMI data give us the first space-based high-resolution imaging spectroscopy of a white-light flare, including continuum, Dop\-pler, and magnetic signatures for the photospheric Fe~{\sc i} line at 6173.34~\AA~and its neighboring continuum.
In the impulsive phase of the flare, a bright white-light kernel appears in each of the two magnetic footpoints.
When the flare occurred, the spectral coverage of the HMI filtergrams (six equidistant samples spanning $\pm$172~m\AA~around nominal line center) encompassed the line core and the blue continuum sufficiently far from the core to eliminate significant Doppler crosstalk in the latter, which is otherwise a possibility for the extreme conditions in a white-light flare.
RHESSI  obtained complete hard X-ray and $\gamma$-ray spectra (this was the first $\gamma$-ray flare of Cycle~24).
The Fe~{\sc i} line appears to be shifted to the blue during the flare but does not go into emission; the contrast is nearly constant across the line profile.
We did not detect a seismic wave from this event.
The HMI data suggest stepwise changes of the line-of-sight magnetic field in the white-light footpoints.

\end{abstract}



\end{opening}

%% In the first two sections, notice the use of the natbib \cite
%% and \inlinecite commands to identify citations.  The citations are


\section{Introduction}

Solar flares are explosive phenomena visible in all regions of the solar atmosphere, and were originally discovered by \inlinecite{1859MNRAs..20...13C} via emission in white light.
Since then, many other manifestations of flares have been discovered in the outer atmosphere and heliosphere, including signatures of acoustic waves penetrating into the solar interior.
According to general consensus, the flare phenomenon corresponds to the sudden 
release of energy stored in the corona via the slow buildup of excess magnetic energy, which ultimately originated via dynamo action within the convective envelope.
Many of the mechanisms remain ill-understood, including the nature of the initial plasma instability
that sets the flare off.
The detection in white light immediately implies a large concentration of the released energy
in the lower atmosphere.
In the upper atmosphere and corona, X-ray and $\gamma$-ray signatures show that the 
energy release has the property of strong particle acceleration, to the extent that major fractions
of the total energy appear to be in electrons above 10~keV and protons above 1~MeV.
These high-energy radiations, and the acceleration of an associated coronal mass ejection (CME), define the ``impulsive phase'' of a flare; other related energy release may take the form of gentler heating.

The original Carrington observation still pre\-sents several open questions.
The white-light flare remains the energetically decisive flare observational signature because most of the flare energy is in the visible and near-UV \cite{2006JGRA..11110S14W,2007ApJ...656.1187F}.
We now think that the visible continuum is enhanced in all flares, but that for the weaker ones the signal is lost in the spatial and temporal brightness fluctuations of the photosphere.
The continuum emission appears in the impulsive phase and is located in the deep solar atmosphere, even apparently reaching the ``opacity minimum'' region of the spectrum near 1.56~$\mu$m \cite{2004ApJ...607L.131X}.
In spite of this supposedly photospheric signal, strong evidence also implicates the chromosphere, since the continuum emission includes clear signatures of recombination radiation \cite{1986lasf.conf..142N,2010arXiv1001.1005H}.
A part of this evidence is the strong association of the white-light continuum with hard X-rays 
\cite{1970SoPh...13..471S,1975SoPh...40..141R,1992PASJ...44L..77H}.
A 10-keV electron has a limited range in a plasma, and cannot reach the photosphere if accelerated in the corona as in the standard thick-target model \cite{1971SoPh...18..489B,1972SoPh...24..414H}.
Accordingly the elucidation of the paths of energy propagation from coronal magnetic storage to the lower atmosphere has great importance.
The Poynting flux could replace the electron beams of the heretofore-standard thick-target model for this purpose \cite{2008ApJ...675.1645F,2009ApJ...695.1151B}, but in any case the strong acceleration of non-thermal electrons above 10~keV remains a requirement.

Observations from the \textit{Helioseismic and Magnetic Imager} (HMI) onboard the {\it Solar Dynamics Observatory} (SDO) give us the first true imaging spectroscopy of flare effects in the photosphere at high spectral and spatial resolution \cite{2010AAS...21630801S} from space.
Previous ground-based observations, typically with slit spectrographs and film readout, have not provided such comprehensive coverage (see \citeauthor{1989SoPh..121..261N} \citeyear{1989SoPh..121..261N} and \citeauthor{2007BCrAO.103...63B} \citeyear{2007BCrAO.103...63B}, for a discussion of this limited material and the conclusions drawn from it).
The new data clearly resolve the profile of the Fe~{\sc i}~line, in each $\approx\ $0.5$''$ pixel and 45-second time step.
The HMI data and RHESSI data for this flare confirm the intimate relationship between flare effects in the lower atmosphere, and high-energy processes revealed by hard X-ray and $\gamma$-ray emissions.
We exploit the new features of HMI to characterize the continuum emission and line-of-sight  Doppler and magnetic properties of the two magnetic footpoints that mark a small but exceptionally impulsive white-light flare.  This study is intended as preparation for future flare observations by SDO and other space-borne and ground-based facilities in Cycle 24.

\begin{table*}
\caption{Flare timeline in HMI data (45-second data frames)}
\centering
\smallskip
\begin{tabular}{l l l l l}
Interval name & Start (UTC) & HMI continuum & HMI Doppler & RHESSI  100~keV\\
\hline
Preflare & 00:54:11 & no excess & no Doppler & no detection \\
Brightening & 00:55:41 & $>$10\% increase & 2~km s$^{-1}$ blue & bright \\
Postflare & 01:00:11 & no excess & complex & no detection \\
\hline
\end{tabular}
\end{table*}
\label{tab:signs}

\section{Observations}

\begin{figure}
\centering
\includegraphics[width=0.8\columnwidth]{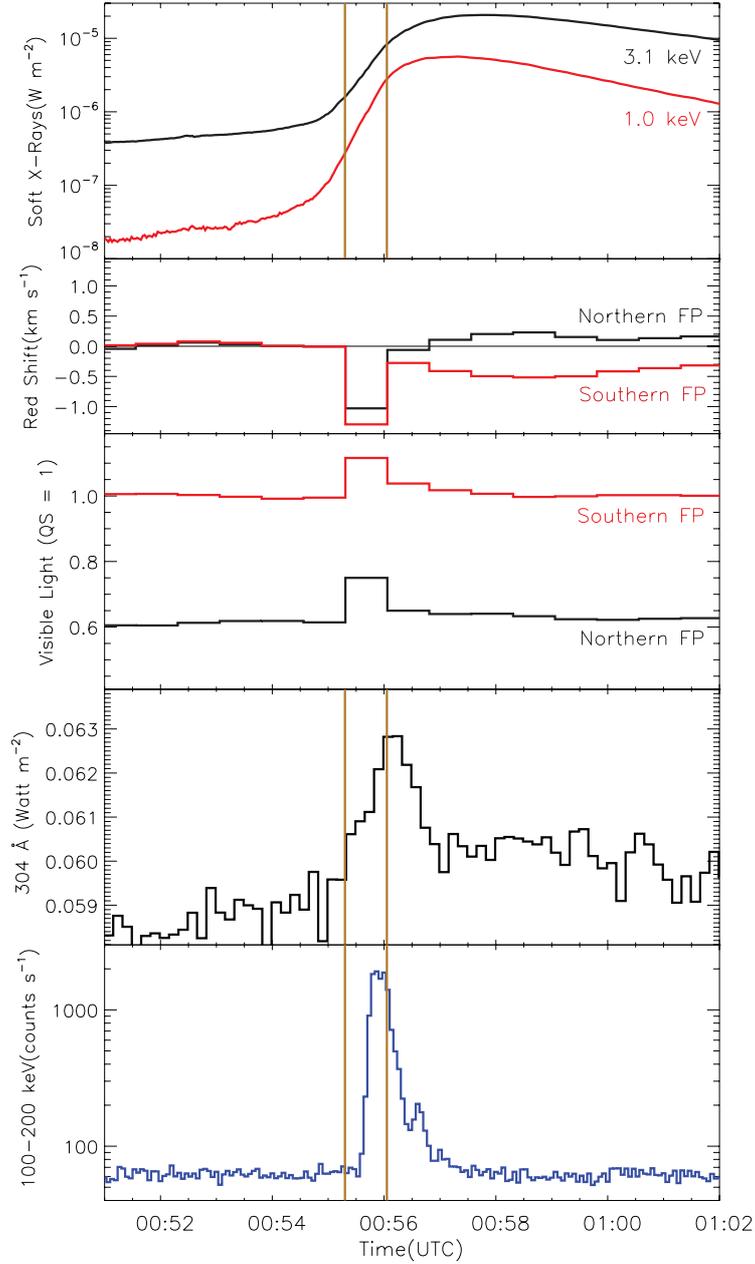}
\caption{Overview of the temporal development of {\tt SOL2010-06-12T00:57}. The five panels show (top to bottom): GOES soft X-rays, HMI NRT (``near real time'') Doppler velocities in the two flaring footpoints (upper line is the northern footpoint), continuum brightnesses in the two regions (``QS = 1'' means that the scale is normalized to the quiet-Sun levels), EVE 304~\AA~data (note suppressed zero), and RHESSI 100\,--\,200~keV hard X-ray fluxes. 
The vertical lines show the Brightening frame (Table~\ref{tab:signs}) beginning and end times.
}
\label{fig:overview} 
\end{figure}

The flare studied was an M2.0~event, hosted by the NOAA active region 11081, located approximately at N22W45 on 12 June 2010.
This flare had a remarkably impulsive hard X-ray light curve, with a duration (half maximum at 50~keV) of only about 25~seconds and $\gamma$-ray emission, unusual for such a weak event \cite{2009ApJ...698L.152S}.
Figure~\ref{fig:overview} brings together some of the time-series information, comparing HMI intensity and Doppler observations  with RHESSI hard X-ray fluxes.
Table~\ref{tab:signs} summarizes the phenomena seen in the relevant HMI intervals and gives them names (Preflare, Brightening, Postflare) for subsequent reference.
The hard X-ray signature mainly matches the Brightening interval (the white-light flare) and does not extend earlier into the previous HMI integration.
The following subsections describe these data in further detail.

\subsection{HMI intensity and Doppler observations}

The continuum observations reported here come from SDO/HMI \cite{2010AAS...21630801S}.
This instrument basically records six-point spectra within $\pm$0.172\AA~of the nominal wavelength
of its target line, a Fe~{\sc i} absorption line at 6173.34~\AA.
The pixel size is $\approx\ $0.5$''$ and the cadence 45~second, dictated by the need to acquire multiple 
exposures with different polarization settings to record nearly simultaneous intensity, Doppler, and magnetic information.
In general, the HMI intensity maps are weighted sums of snapshot filtergrams over a range of times before and after the times that the respective maps represent.  

For the regular helioseismic applications, the weighting function extends to a range of 135~seconds before and after the ``record time'' represented by
the snapshot and has negative sidelobes in the range $\pm$45--90~seconds (see Figure~\ref{fig:artifact}).  Applied to a highly impulsive flare of short duration,
such as {\tt SOL2010-06-12-T00:57}, the sidelobes can produce significant artifacts, specifically ringing before and after the onset of the flare
that appears as a spurious ``black-light'' precursor (\opencite{1990A&A...233..577H}, see Appendix). Actual negative flare phenomena are often observed in stellar flares \cite{1982ApJ...252L..39G} and would have interesting diagnostic uses if detected in solar flares. Because of this interesting possibility we describe the effects of the interpolation function in the Appendix.

\begin{figure}
\centering
\includegraphics[width=\columnwidth]{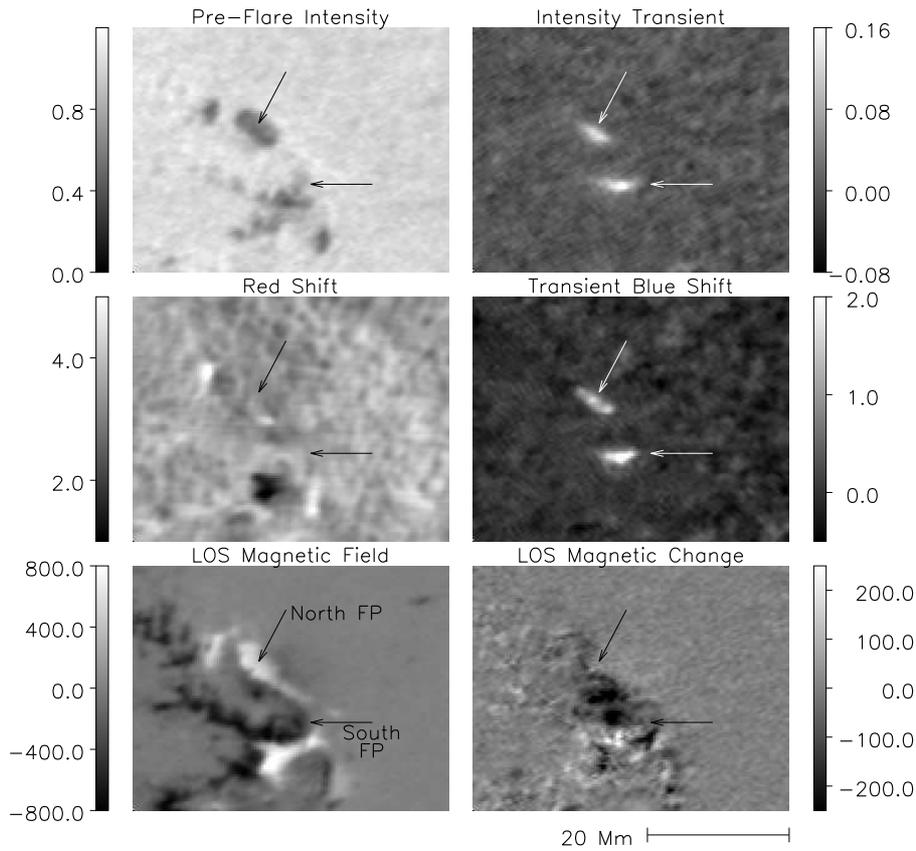}
\caption{HMI NRT (``near real time'') images: upper left, the Preflare continuum image; upper right, the Brightening continuum minus the Preflare continuum; middle left, Doppler maps average over 55~minutes preflare interval;  middle right, difference between the Brightening Doppler map and immediately prior to the pre-flare map; lower left, the Preflare line-of-sight field; and lower right, the Postflare line-of-sight field minus the Preflare line-of-sight field (see Table~\ref{tab:signs} for times).
The arrows locate the footpoint sources.
The grayscale shows fractions of quiet-Sun intensity, km s$^{-1}$, and Gauss for the three rows.
} 
\label{fig:fig2a}
\end{figure}

\begin{figure}
\centering
\includegraphics[width=0.8\columnwidth]{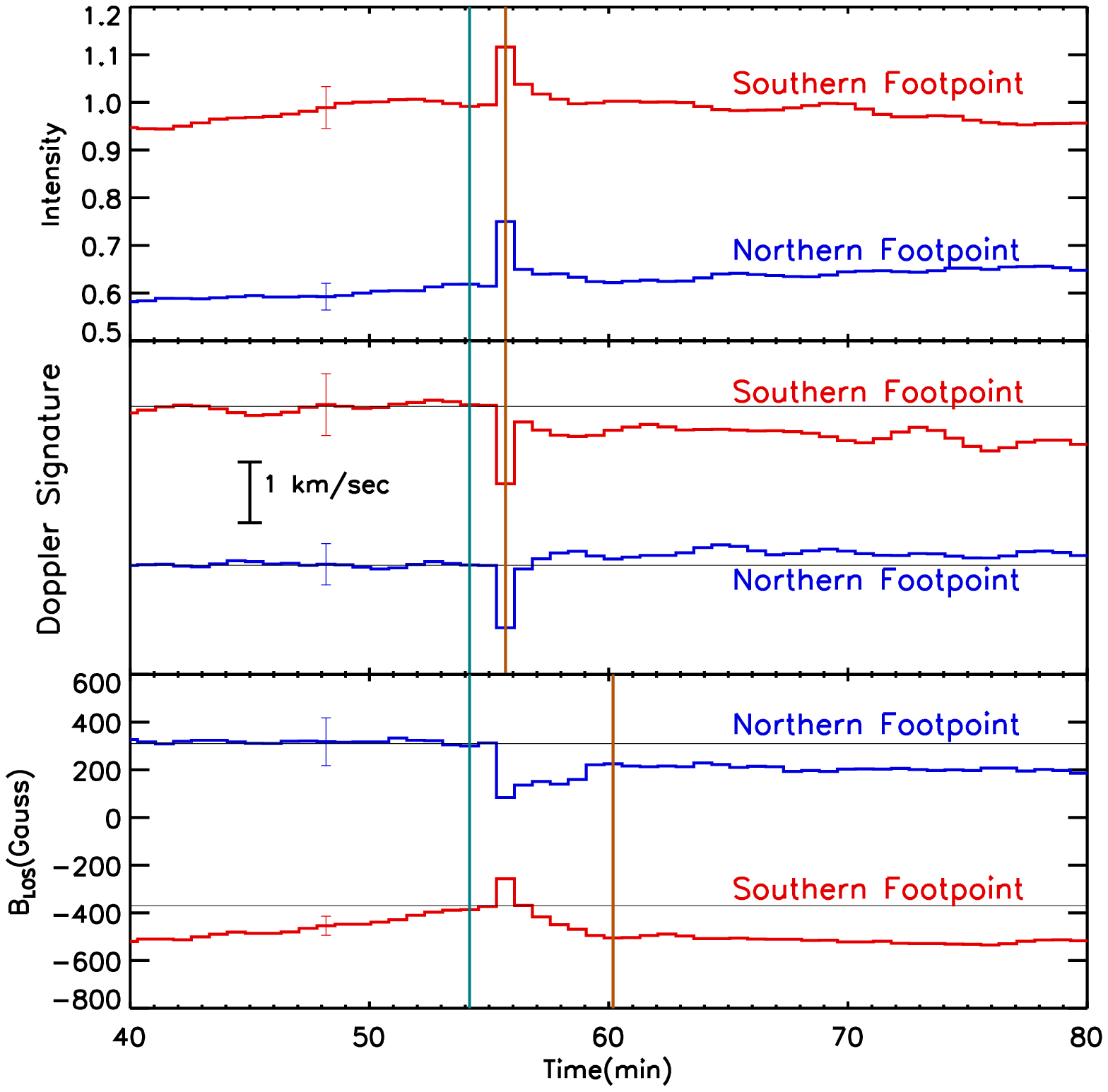}
\caption{NRT time series of intensity, Doppler velocity, and mean magnetic flux density for the two flaring footpoints. The turquoise vertical lines represents the frame before the onset of the flare.  The brown vertical line represents flare maximum in the top two panels and the Post-flare reference in Table~\ref{tab:signs} in the bottom panel. 
Error bars represent $\pm$~ten times the pre-flare rms consecutive-sample differences in the respective intensity, Doppler, and magnetic values over a 10-minute period.
In each panel except the bottom the southern footpoint is the upper line.
}
\label{fig:time_series} 
\end{figure}

To avoid the pre-flare artefacts introduced by the extended negative weightings in the normal helioseismic database, the intensity and Doppler maps analyzed here are taken from the ``near-real-time'' (NRT) database.  
These are computed by simple linear interpolation, with positive weightings, between pairs of filtergrams within 45~seconds of the times represented by respective maps.  
Figure~\ref{fig:fig2a} shows NRT maps of intensity, Doppler and magnetic signatures of the flare immediately before and during the onset of impulsive white-light emission.
Figure~\ref{fig:time_series} shows the time series of intensity and Doppler signal for the two white-light-flaring regions, which we identify as footpoints of a coronal-loop structure.

\begin{figure}
\centering
\includegraphics[width=0.6\columnwidth]{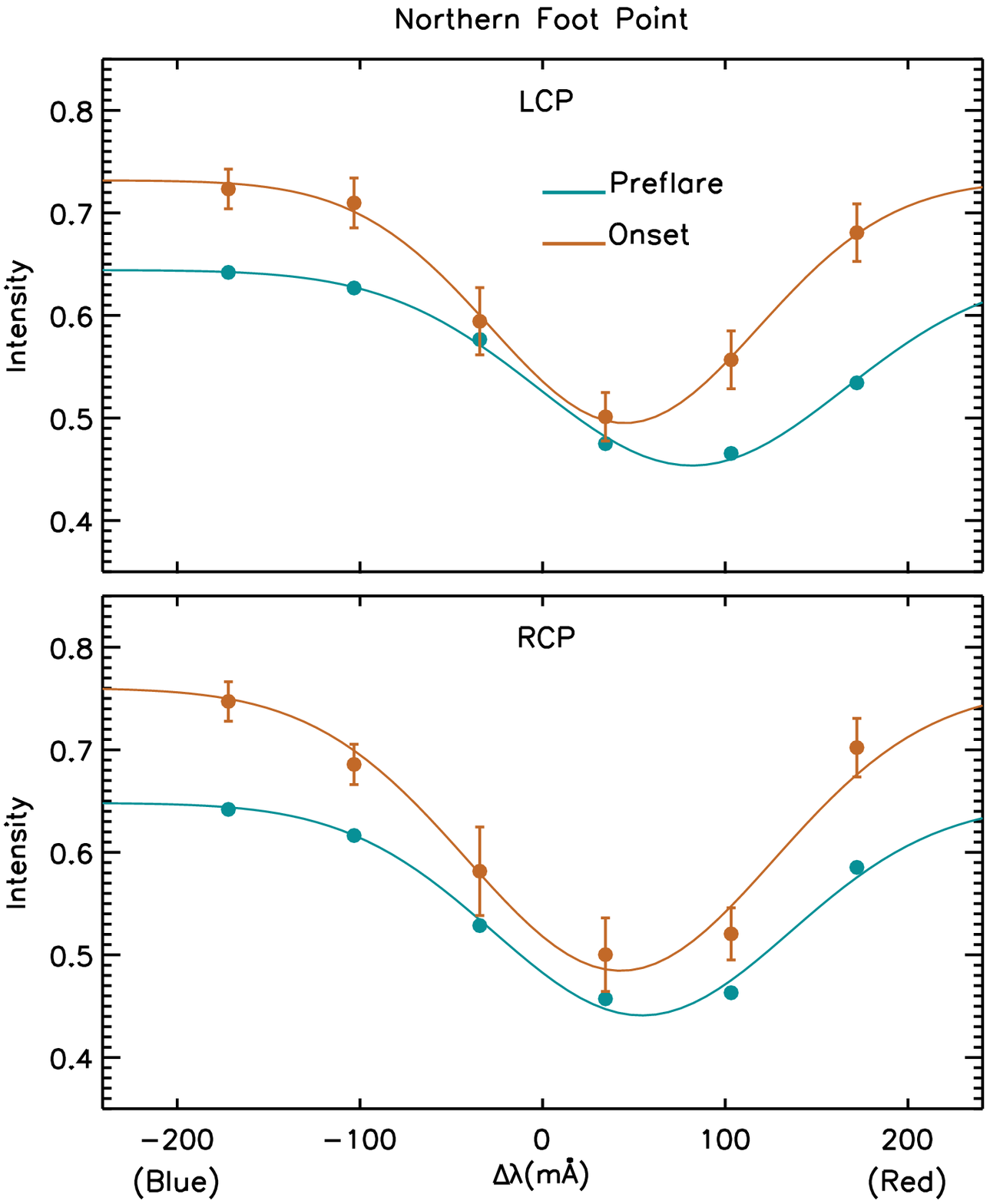}
\caption{Six-point spectra obtained by HMI in the northern flare region in both circular polarizations, with four-parameter Gaussian fits. 
The plots show both senses of circular polarization for the summed pixels of the northern footpoint,
with the upper line in each from the Brightening frame and the lower line the Preflare frame.
Error bars represent $\pm$~ten times the pre-flare rms consecutive-sample differences in the respective intensity over a ten-minute period.
}
\label{fig:spectra} 
\end{figure}

In the following discussion we deal only with NRT data and single filtergrams, and study its high-resolution spectral information at the standard sampling
interval of 45~seconds for the line-of-sight field data. Figure~\ref{fig:spectra} shows the filtergram intensities and spectral fits for the northern flare region, in both circular polarizations.
The spectra show that the Fe~{\sc i}~line moves significantly to the blue in the Brightening frame, in both footpoints.
The four-parameter Gaussian fits show little or no tendency for simple line broadening.
Because of the motions of the spacecraft at the time of the observation, and for other reasons, the spectral line generally does not lie exactly in the center of the spectrum.
In this case the blue continuum is seen slightly better, which gives a fortuitous advantage in identifying the continuum.
During the event, the line moves bodily, retaining its approximate pre-flare width; this suggests a degree of uniformity across the flare area.
If the profile resulted from the average of brief impulses in small areas, one would expect line broadening.

Table~\ref{tab:fits} gives the parameters derived from the Gaussian fits to the mean line profiles from the north and south footpoints, comparing the Preflare and Brightening frames.
The uncertainties in the fitted parameters can be judged by the quality of the fits shown in Figure~\ref{fig:spectra}.

\begin{table}[h!]
\caption{Fit Parameters}
\smallskip
\begin{tabular}{l l l l l l r}
 &  Frame &   & Continuum & Line & Width  & Center \\
 &  &   &  Intensity     & Strength & (m\AA)    & (m\AA) \\
 \hline
North & Preflare & LCP & 0.64     &   0.30    &     119      &    82  \\
          &      & RCP &    0.65   &    0.32   &      115    &      54  \\
 &Brightening      & LCP  &    0.73  &       0.32 &        104    &      45  \\
  &               & RCP    &  0.76   &      0.36    &     119     &     43  \\
\hline
South & Preflare & LCP  &    0.96    &     0.30     &    109      &    72  \\
        &  &  RCP &     0.95  &      0.27    &     107     &    113  \\
& Brightening  &   LCP  &    1.09    &     0.30     &    101    &      41 \\
 &       &    RCP &     1.10    &     0.25    &     116      &    60 \\
\hline
\end{tabular}
\label{tab:fits}
\end{table}

\subsection{RHESSI Hard X-rays}

As shown in Figure~\ref{fig:overview}, the hard X-ray/$\gamma$-ray time profile of this event was 
simple, with the appearance of a single spike with a duration of about six~RHESSI four-second integrations, split between the Brightening and Postflare frames but mainly in the former.
Because of a special observing program for RHESSI at the time of this flare, we do not have images for the hard X-rays or $\gamma$-rays.
We have confirmed the timing of the hard X-ray burst by comparison with \textit{Fermi} burst data (B.~Dennis, R.~Schwartz, and K.~Tolbert, personal communication 2010), finding agreement to within the basic four-second time resolution of RHESSI.
Both RHESSI and \textit{Fermi} detected this flare in $\gamma$-rays as well -- a somewhat unusual property for an M2-class event \cite{2009ApJ...698L.152S}.
The hard X-ray burst is well-defined and has a sharp onset at $\approx\ $100~keV.
At the resolution of the HMI observations, the hard X-rays and visible continuum occur simultaneously, as expected from earlier observations \cite{1975SoPh...40..141R,1992PASJ...44L..77H}

\subsection{Magnetic Fields}

\begin{figure}[ht]
\centering
\includegraphics[width=0.95 \columnwidth]{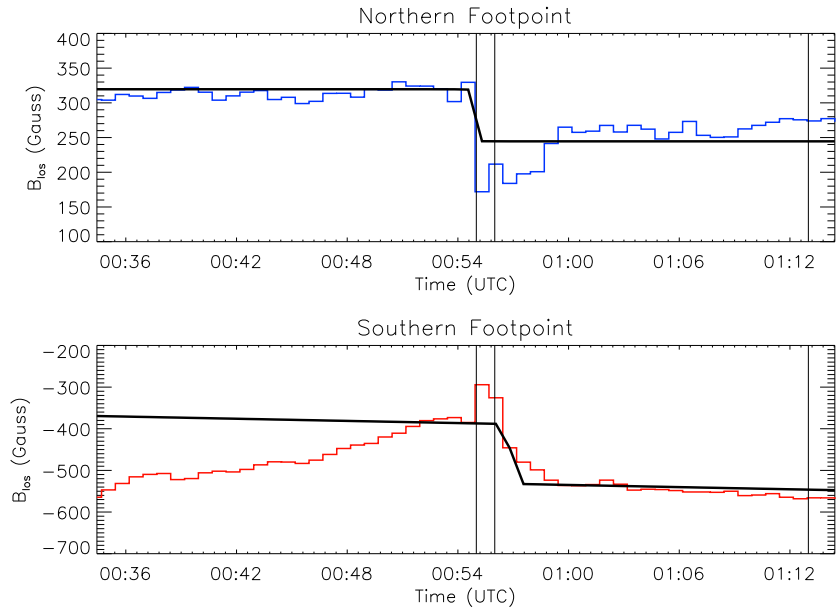}
\caption{Demonstration of a stepwise change in the line-of-sight field components of the field in the southern region (\textit{lower}) and the northern region (\textit{upper}).
The black (thick) curves are mathematical fits to the stepwise variations following the method of Sudol and Harvey (2005).
These fits use $\pm$15~minutes of data around the flare. 
The vertical lines show the GOES start, maximum, and end time. 
}
\label{fig:sudol} 
\end{figure}

Figure~\ref{fig:sudol} also reveals variations consistent with the occurrence of stepwise variations of the line-of-sight magnetic field \cite{1994ApJ...424..436W,1999SoPh..190..459K} as shown in detail for many X-class flares by \cite{2005ApJ...635..647S}.
This is one of the first flares for which HMI data could be used for this purpose, and it is clear that the signal-to-noise level is adequate to study such effects for an event of this magnitude.
Note that the signs of the before-to-after changes in the line-of-sight magnetic field are the same, even though the line-of-sight components of the magnetic fields themselves are of opposite sign.
This differs from the expectation from the \inlinecite{2005ApJ...622..722L} observations of penumbral changes, which suggest a more symmetrical loop contraction.
The fits place the time of the step close to the impulsive phase (as estimated from the GOES times shown in Figure~\ref{fig:sudol}).
This would be expected because of the major energy requirements at this time (\textit{e.g.} \opencite{2000ApJ...531L..75H}) and is typical of the Sudol and Harvey results.

The Doppler signature directly detected in the line profile reveals that the flare caused an impulse at the (photospheric) altitude of the line-formation region.
As shown by the spectra (Figure~\ref{fig:spectra}), this impulse is a strong blue shift, and this is evident at both footpoint sources.
\inlinecite{koso2006K} describes ``up and down motions'' and ``strong down flows'' to two of the 
seismic source regions in his survey of MDI detections, which is different from what
we observe.
We note that this flare did not produce a detectable seismic signature, as predicted by \inlinecite{1972ApJ...176..833W} and observed by \inlinecite{1998Natur.393..317K} and others (\textit{e.g.}, \opencite{2005ApJ...630.1168D}).
A detailed analysis of the magnitude of the impulse and the upper limits to seismic wave energy in the solar interior is beyond the scope of this introductory paper, but the data clearly would be an excellent starting point for modeling that will help to clarify the energy and momentum transfer in seismic wave excitation.

\section{Conclusions}

The HMI imaging spectroscopy of this flare resulted in maps of the line profile of the photospheric line Fe~{\sc i} at 6173.34\AA.
The information includes line width, depth, Doppler shift, and line-of-sight magnetic field via observations in two polarizations.
They also record the neighboring continuum near 6173.34~\AA.
This capability greatly improves our understanding of the lower solar atmosphere during flares,
since the full information of the line profile can be interpreted in terms of the physical conditions
there.
The full imaging spectroscopy allows us to identify the continuum brightenings with the footpoint regions of coronal loop structures via reference to AIA images.
A fuller analysis of these data (and those of EVE) is outside the scope of this paper.

For the flare studied here we find that the line shifts in wavelength, but remains in absorption.
The intensity in the core of the line appears to have approximately the same flare excess as the nearby continuum (cf. \opencite{2010ApJ...722.1514P}).
We also find strong blueshifts of the line at both footpoints.
Both of these findings are significant of multiple possibilities.
Backwarming \cite{1989SoPh..124..303M} due to the observed continuum should raise the photospheric temperature, which could, in principle, change both the strength and width of the line. The observation of numerous other white-light flares with HMI should restrict these possibilities; we cannot at present rule out the possibility that the blueshifts could be an artifact of the sampling sequence.

The transient blueshift could signify a photospheric medium moving toward the spacecraft, shifting the absorption line accordingly.
Alternatively, it could be the result of red-shifted component of line emission, from a down-flowing heated chromosphere\footnote{The latter would be consistent with red-shifted H$\alpha$ emission in reaction to chromospheric ablation at higher altitudes \cite{2009ApJ...699..968M} and in Na\,D$_1$-line emission seen by \inlinecite{2005ApJ...630.1168D}.  
The chromospheric emission would have to be insufficiently strong to drive the photospheric line into emission, but the resultant of the superposition would be a blue-shifted absorption profile.}.  
Detailed radiative-transfer modeling of appropriate scenarios is needed to address this question. 
It is also tantalizing is that the transient blue shift is so apparently at odds with the transient red (or mixed) Doppler shifts seen in other observations \cite{koso2006K,2007AdSpR..40.1921B}.

For a magnetic region as far limbward as AR11081 at the time of {\tt SOL-2010-06-12T00:57}, it is important to consider horizontal motion of the medium.  
This could either be motion driven by a magnetic jerk or magnetic deflection of motion that would have been vertical except for a strong, inclined magnetic field.  A basic control question for this hypothesis, then, is whether flares from near disk center show the transient red shifts generally seen in MDI observations, as Doppler observations of these should be insensitive to horizontal motion.

A final consideration, given the suddenness and short duration of the HXR profile in {\tt SOL2010-06-12T00:57}, is the possibility of a blueshift artificially caused by aliasing.
The time separation of red and blue spectroheliograms, in the presence of a varying line intensity, could result in a spurious Doppler signal.
This possibility can be controlled by comparisons with GONG observations of {\tt SOL2010-06-12-T00:57}, in which temporal aliasing is greatly
reduced in integrations of respective intensity and Doppler signatures  for the full duration represented by the record time.  This comparison is 
being undertaken in a study in progress that benefits from new analysis techniques that allow us to compensate for noise introduced by variations 
in atmospheric seeing quality \cite{2008SoPh..251..627L}.

The HMI observations are consistent with the idea that the flare emission at this wavelength has a substantial component of Paschen continuum from hydrogen recombination at higher altitudes, a conclusion also consistent with observations of the Balmer and even Paschen continuum edges in the spectra of other white-light flares \cite{1989SoPh..121..261N}.
To understand these results quantitatively will require modeling beyond the scope of this paper, and of course to draw any general conclusion would require the observation of other flares in this manner as well.

The HMI observations clearly point to the need for higher temporal resolution along with good spatial resolution in white light.
Ground-based observations (\textit{e.g.,} \opencite{2008ApJ...688L.119J}) could extend this work, but we also expect many more interesting flare observations from the powerful instrumentation on SDO.
We expect that comparable HMI data will become available for many other flares and that the modeling of the lower solar atmosphere will solve several outstanding problems of interpretation, including the nature of the impulses responsible for sunquakes \cite{1998Natur.393..317K}.

\bigskip\noindent{\bf Acknowledgments:}
The Berkeley group was supported by NASA under contract NAS 5-98033 for RHESSI, and the
SDO/HMI by contract NAS5-02139 to Stanford University.
We thank the other members of the SDO/HMI team for special help during a very busy time, and for building such a fine instrument.
These data have been used courtesy of NASA/SDO and the AIA, EVE, and HMI science teams.
The author list of this paper consists of those persons actively involved in actually writing the paper, rather than making it possible, and after the first author's name the order has been randomized.

\appendix 
\label{sec:appendix}

\begin{figure}[htb]
\centering
\includegraphics[angle=90,width=0.5\columnwidth]{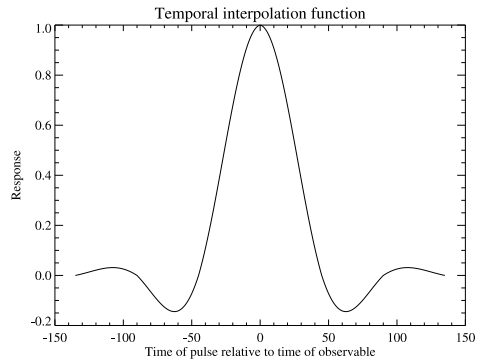}
\caption{Plot of the function specifying the interpolation coefficients applied to filtergrams for computation of Doppler, intensity, and line-of-sight magnetic maps in HMI time series.  Note the up-to-15\% reversal in the value of the function in the intervals $\pm$45--90~seconds.} 
\label{fig:filter}
\end{figure}

\begin{figure}
\centering
\includegraphics[width=\columnwidth]{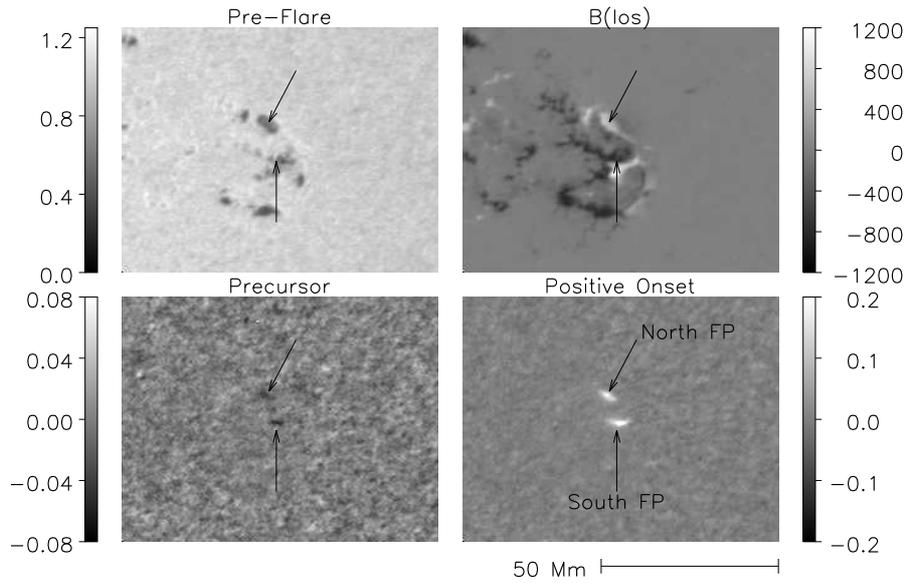}
\caption{Intensity artifact prior to a white-light flare in helioseismically interpolated intensity maps.  Upper left frame shows pre-flare intensity.  Upper right frame shows the pre-flare line-of-sight magnetic field.  Bottom row shows consecutive intensity-difference images.  Bottom left shows the pre-flare difference; bottom right shows the white-light-flare difference.  Note the apparent reduction in intensity in the flare footpoints in the lower left frame.} 
\label{fig:artifact}
\end{figure}

The HMI Dopplergrams and intensities used for standard helioseismic applications of the data consist of a spatial-temporal interpolation, using 12~filtergrams at six different wavelengths and two polarization states, for each observable. 
These interpolations, in conjunction with the scanning of the Fe~{\sc i} line, are devised to minimize the effects of aliasing in \textit{p}-mode recognition.  
The standard interpolation for this purpose includes contributions from 135 seconds before and after the time assigned to a Dopplergram, with coefficients that are negative in the ranges $\pm$45\,--\,90~seconds (see Figure~\ref{fig:filter}).
Because of this, the response of the standard Dopplergrams to a sufficiently sharp white-light flare can be an apparent reduction in intensity preceding the flare, \textit{i.e.}, an apparent ``black-light flare'' preceding the white-light excess (see Figure~\ref{fig:artifact}). 

\bibliographystyle{spr-mp-sola}
\bibliography{blf}

%\end{twocolums}
\end{article} 
\end{document}